\documentclass[a4paper]{PoS}

\usepackage{amsmath}

\newcommand{\eVdist}{\kern-0.06667em}
\newcommand{\mev}{{\,\text{Me}\eVdist\text{V\/}}}
\newcommand{\gev}{{\,\text{Ge}\eVdist\text{V\/}}}
\newcommand{\ket}[1]{\left|\rule{0pt}{10pt} #1 \right>}
\newcommand{\bra}[1]{\left< #1 \rule{0pt}{10pt}\right|}
\newcommand{\bret}[1]{\left< #1 \right>}
\newcommand{\op}[1]{\mathcal{#1}}
\newcommand{\I}{\mathrm{i}}
\newcommand{\di}{\mathrm{d}}
\newcommand{\e}{{e}} 
\newcommand{\lrd}{\raisebox{0.09em}{$
\stackrel{\raisebox{-0.03em}{$\scriptstyle\leftrightarrow$}}{D}$}{}}
\newcommand{\msb}{\text{$\overline{\text{MS}}$}}
\newcommand{\Tr}{\mathrm{Tr}}

\title{Parton Distribution Amplitudes and Non-Perturbative Renormalisation%
  \begin{picture}(0,0)
    \put(129,80){\makebox(0,0)[r]{\normalfont\normalsize{SHEP-08-28}}}
  \end{picture}
}

\ShortTitle{Parton Distribution Amplitudes and Non-Perturbative Renormalisation}

\author{P.A.\ Boyle$^a$, \speaker{D.\ Br\"ommel}$^b$, M.A.\ Donnellan$^{b,c}$,
  J.M.\ Flynn$^b$, A.\ J\"uttner$^d$ and C.T.\ Sachrajda$^b$\\
  \llap{$^a$} School of Physics and Astronomy, University of Edinburgh,
  Edinburgh EH9 3JH, UK\\
  \llap{$^b$} School of Physics and Astronomy, University of Southampton,
  Southampton SO17 1BJ, UK\\
  \llap{$^c$} Deutsches Elektronen-Synchrotron DESY, 15738
  Zeuthen, Germany\\
  \llap{$^d$} Institut f\"ur Kernphysik, Johannes Gutenberg-Universit\"at Mainz,
  55099 Mainz, Germany\\
  E-mail: \email{d.broemmel@phys.soton.ac.uk}}

\author{RBC/UKQCD Collaborations}

\abstract{ We present results for the first two moments of the light-cone
  distribution amplitudes of the $\pi$ and $K$ pseudo-scalar mesons and of the
  $\rho$, $K^{\ast}$ and $\phi$ vector mesons. The calculations are performed on
  the RBC/UKQCD collaborations' ensembles generated with the Iwasaki gauge
  action and with 2+1 flavours of domain wall fermions. In addition we also
  provide some results on the necessary non-perturbative renormalisation which we
  perform using the Rome-Southampton method. We discuss the benefits of the
  momentum source approach such as much smaller statistical errors and the
  possibility to see effects of the discretisation.  }

\FullConference{The XXVI International Symposium on Lattice Field Theory\\
  July 14-19 2008\\
  Williamsburg, Virginia, USA}

\begin{document}

\section{Introduction}

The first part of these proceedings will give an update on our results for the
lowest moments of the leading twist meson distribution amplitudes (DAs)
\cite{Boyle:2006pw,Donnellan:2007xr}. Distribution amplitudes contain
non-perturbative QCD effects that appear e.g.\ in hard exclusive processes but
are universal hadronic properties and so do not depend on the process itself.
They are important for form factors at large $q^2$ or $B$-decays and can be
related to the Bethe-Salpeter wave function. Here we provide results for the
pseudo-scalar DAs of $K$ and $\pi$ as well as for the vector DAs of $K^*$,
$\rho$ and $\phi$. More details on these calculations by the RBC and UKQCD
collaborations will be presented in a forthcoming paper.

The second part is devoted to the non-perturbative renormalisation of quark
bilinears with and without derivatives. Renormalisation is necessary to obtain
physical results for e.g.\ the matrix elements of moments of DAs or decay
constants. We follow the Rome-Southampton method \cite{Martinelli:1994ty} with
momentum sources \cite{Gockeler:1998ye} to calculate the renormalisation
constants. The first is aimed at reducing uncertainties in the perturbative
renormalisation while the latter is to improve upon the statistical errors from
the standard point source approach.

The calculations are done on lattices with $24^3\times 64\times 16$ and
$16^3\times 32\times 16$ points at a lattice spacing of $a^{-1}=1.729(28)\gev$.
The ensembles have been generated by the RBC/UKQCD collaborations using
$N_\text{f}=2+1$ domain wall fermions with an Iwasaki gauge action. Our light
quark masses range from $am_\text{q}=0.005$ to $0.03$ corresponding to pion
masses from $331\mev$ to $672\mev$. The strange quark mass is kept fixed at
$am_\text{s}=0.04$. Details on the ensembles have been reported in
\cite{Allton:2007hx, Allton:2008pn}.

\section{Meson distribution amplitudes}

The meson DAs are defined as non-local matrix elements on the light-cone. The
leading twist pseudo-scalar and (longitudinal) vector DAs
are
\begin{align}
  \label{eq:defDA}
  \bra{0} \bar{\psi}_{F_1}(z) \gamma_\mu \gamma_5 \op{U}(z,-z) \psi_{F_2}(-z)
    \ket{\Pi^+(p)}\Bigr|_{z^2=0}
  &= \I f_\Pi p_\mu \int_{-1}^1 \di{\xi} \e^{\I \xi p\cdot z} \phi_\Pi(\xi,\mu),\\
  \bra{0} \bar{\psi}_{F_1}(z) \gamma_\mu \op{U}(z,-z) \psi_{F_2}(-z)
    \ket{V(p,\lambda)}\Bigr|_{z^2=0}
  &= f_V m_V p_\mu \frac{\epsilon^*(\lambda)\cdot z}{p\cdot z}
    \int_{-1}^1 \di{\xi} \e^{\I \xi p\cdot z} \phi_V^\parallel(\xi,\mu).
\end{align}
Here $\xi=2x-1$ with $x$ and $1-x$ the momentum fractions of the two quarks. The
Wilson line $\op{U}$ ensures gauge invariance and the quark flavours $F_i$ are
chosen to match the pseudo-scalar and vector mesons. All of the DAs
$\phi(\xi,\mu)$ are normalised to unity when integrated over $\xi\in [-1,1]$. On
the lattice we only access moments of the DAs,
\begin{equation}
  \label{eq:momDA}
  \bret{\xi^n}(\mu) = \int \di{\xi} \xi^n \phi(\xi,\mu),
\end{equation}
which appear in matrix elements of local operators with $n$ derivatives, see
e.g.\ \cite{Donnellan:2007xr}. We extract the bare values for
$\bret{\xi^n}^\text{bare}\; (n=1,2)$ from the lattice using ratios of two-point
functions. To give one example, let us consider the second moment of the
(longitudinal) vector meson DA, $\bret{\xi^2}^{\parallel ,\text{bare}}(\mu)$.
For large Euclidean times $t$ and $T-t$, the correlation functions give for a
ratio like
\begin{multline}
  \label{eq:ratio}
  \frac{\sum_{\vec{x}} \e^{\I \vec{p} \vec{x} }\bra{0}
    \op{O}_{\{\rho\mu\nu\}}(\vec{x},t) V_\sigma^\dagger(0) \ket{0}}
  {\frac{1}{3}\sum_i\sum_{\vec{x}} \e^{\I \vec{p} \vec{x}}
    \bra{0} \op{O}_{i}(\vec{x},t) V^\dagger_i(0)\ket{0}} \,\longrightarrow\,
  -\I \bret{\xi^2}^{\parallel,\text{bare}} \tanh\left((t-T/2)E_V\right)\\
  \times \frac{1}{3}
  \biggl(-g_{\rho\sigma} p_\mu p_\nu - g_{\mu\sigma}p_\rho p_\nu
  -g_{\nu\sigma} p_\rho p_\mu + \frac{3p_\rho p_\mu p_\nu p_\sigma}{m_V^2}\biggr).
\end{multline}
Here $E_V$ and $m_V$ are the energy and mass of the vector meson with
interpolating field $V$. The operator with $n$ derivatives is given by
\begin{equation}
  \label{eq:op}
  \op{O}_{\{\mu\mu_1\dots\mu_n\}}(\vec{x},t) = \bar{\psi}_{F_1}(\vec{x},t)
  \gamma_{\{\mu} \lrd_{\mu_1} \dots \lrd_{\mu_n\}} \psi_{F_2}(\vec{x},t),
\end{equation}
where the $\{\dots\}$ denote symmetrisation of the indices and subtraction of traces.
Note that by choosing directions such as e.g.\ $\mu=\sigma=2$, $\nu=4$ and $\rho=1$
one unit of momentum $p_1 \ne 0$ is enough to extract $\bret{\xi^2}^{\parallel
  ,\text{bare}}(\mu)$ from Eq.~\eqref{eq:ratio}.

\begin{figure}[!t]
  \parbox{.5\textwidth}{\centering\tiny\input{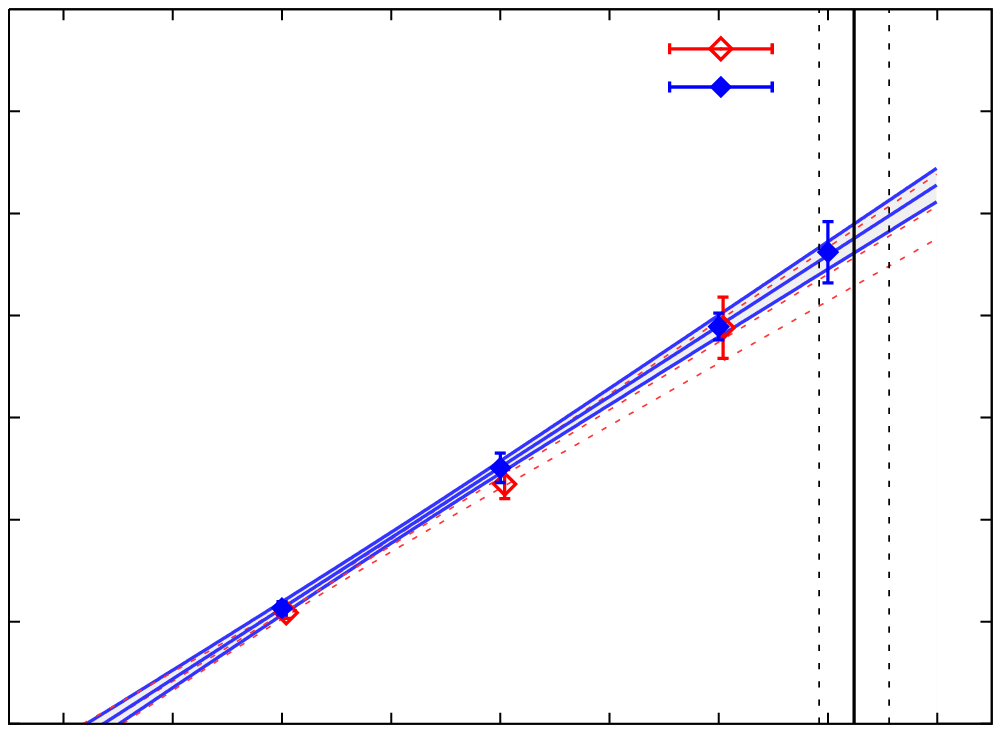}}
  \parbox{.5\textwidth}{\centering\tiny\input{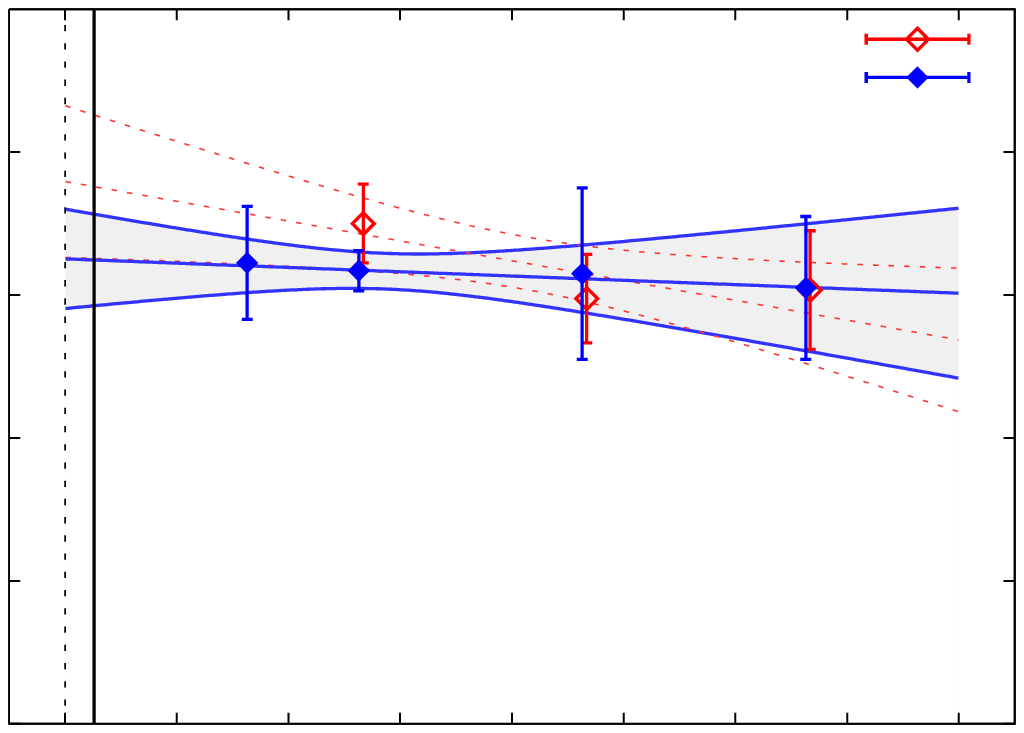}}
  \parbox{.5\textwidth}{\centering\tiny\input{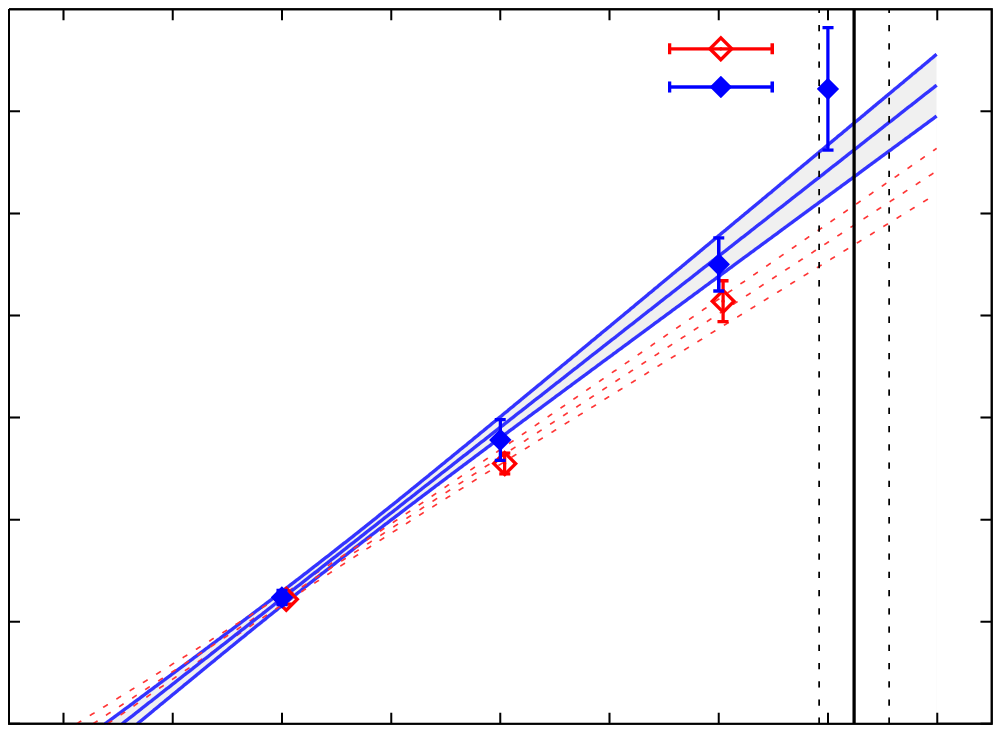}}
  \parbox{.5\textwidth}{\centering\tiny\input{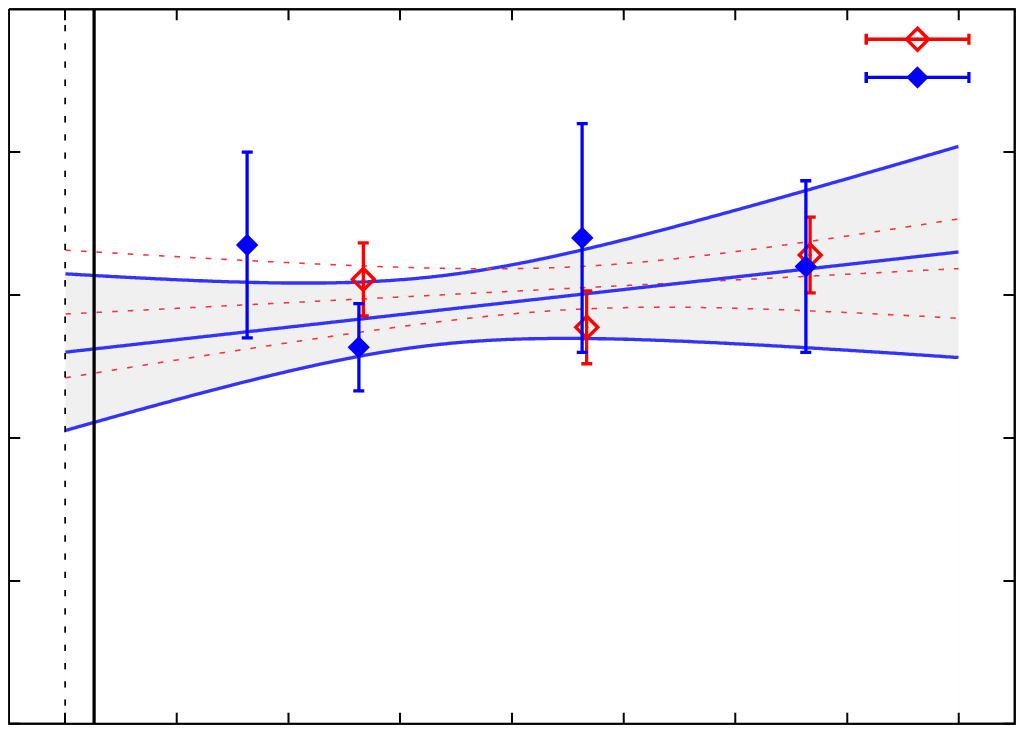}}
  \parbox{.5\textwidth}{\centering\tiny\input{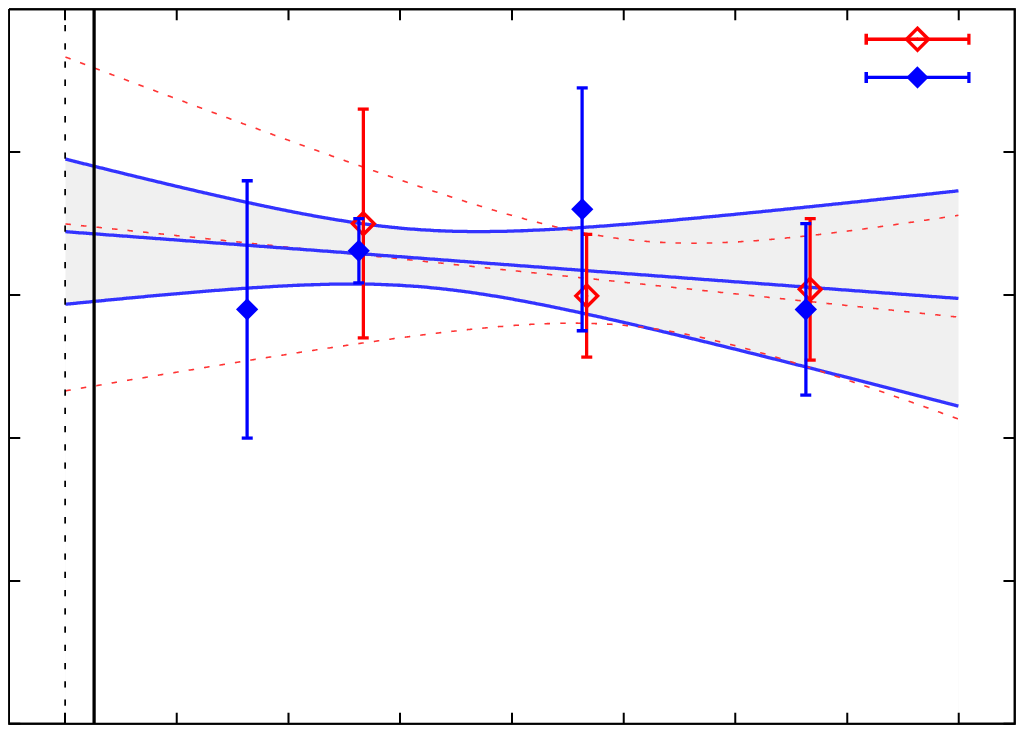}}
  \parbox{.5\textwidth}{\centering\tiny\input{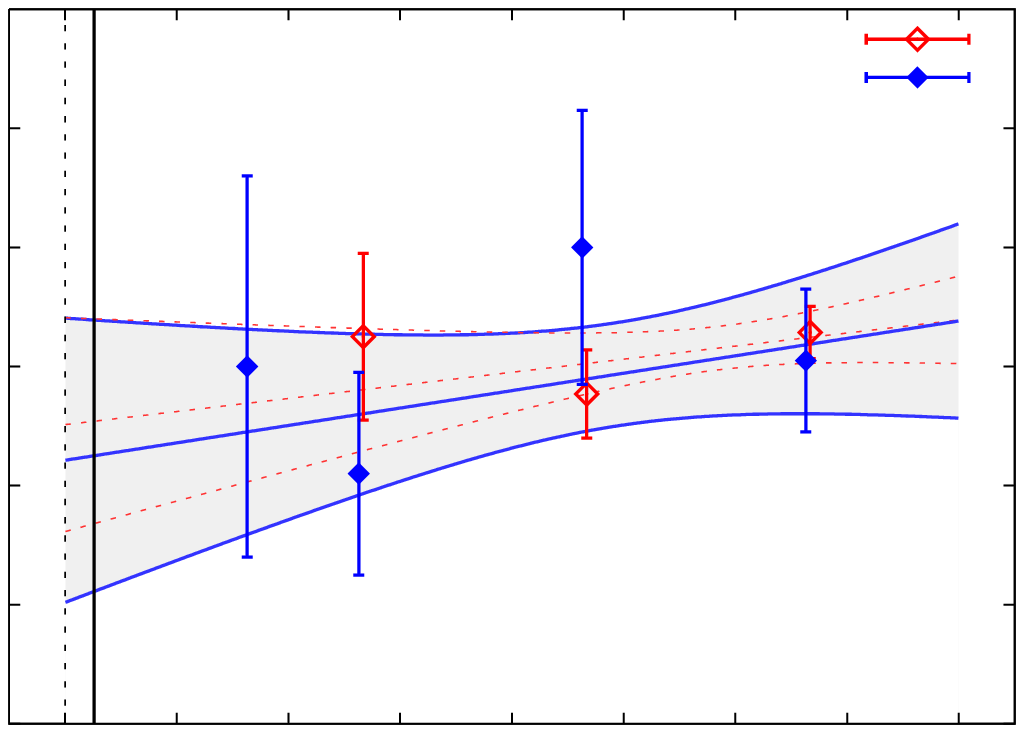}}
  \parbox{.5\textwidth}{\centering\tiny\input{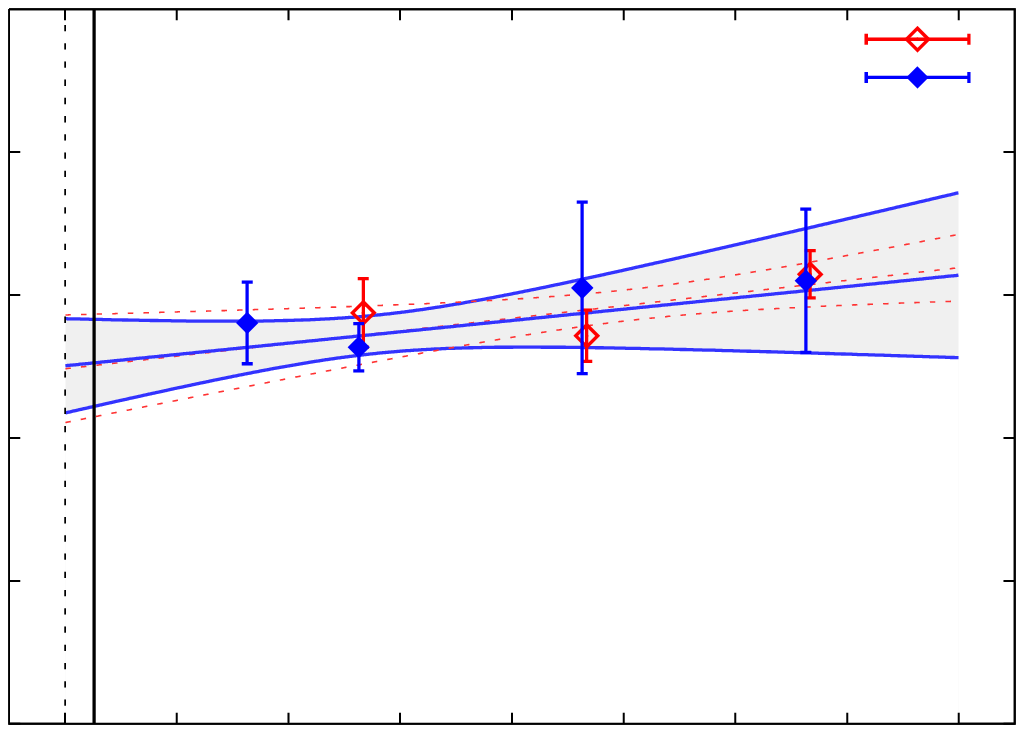}}
  \parbox{.5\textwidth}{\caption{\label{fig:DAs}Linear extrapolations of the
      various moments of the DAs for pseudo-scalar and vector mesons against the
      mass difference $m_\text{s} - m_\text{q}$ or the quark mass. Shown are the
      bare results for $K$, $K^*$, $\pi$, $\rho$ and $\phi$ for both lattice
      sizes. The solid vertical lines mark the physical point with the
      uncertainty in $m_\text{s}$ in case of the lowest moments of $K$ and
      $K^*$.}}
\end{figure}
The results of fits to ratios like Eq.~\eqref{eq:ratio} are plotted for both
lattice sizes in Fig.~\ref{fig:DAs}. Also shown are linear extrapolations of
these bare results to the chiral limit along with their error bands. For the
lowest moment of the kaon DA such a linear extrapolation in $(m_\text{s} -
m_\text{q})$ is predicted by NLO chiral perturbation theory \cite{Chen:2003fp}.
Note that we have to account for a slightly wrong strange quark mass of
$am_\text{s} = 0.04$ in our simulation instead of the physical $am_\text{s} =
0.0343(16)$ \cite{Allton:2008pn}. In case of the lowest moment for $K$ and
$K^*$, this is done \cite{Donnellan:2007xr} by extrapolating in $(m_\text{s} -
m_\text{q})$ where we can simply enter the correct physical quark masses. For
the remaining second moments, a correction in $m_\text{s}$ could be obtained by
comparing the results for mesons with and without strange quarks.  However,
within errors these differences are small enough to neglect any corrections.
Fig.~\ref{fig:DAs} also shows that finite size effects are small and not
  significant within errors. Finally, the preliminary results for our
$\bret{\xi^n}\; (n=1,2)$ at the physical point are given in Tab.~\ref{tab:DAs},
along with results by other groups where available. Our values are quoted using
perturbative renormalisation to the \msb\/ scheme at a scale of $\mu = 2\gev$.
We have already pointed out the clearly observable $SU(3)$-breaking effects for
the kaon DA in \cite{Boyle:2006pw}. The same now holds for the $K^*$.
\begin{table}[!b]
  \setlength{\belowcaptionskip}{\abovecaptionskip}
  \setlength{\abovecaptionskip}{0pt}
  \centering
  \caption{\label{tab:DAs}Our preliminary results at the physical point and both
    our lattice sizes, in \msb\/ at \protect{$\mu = 2\gev$} using perturbative
    renormalisation. Also included are results from the given references (at
    $\mu=2\gev$).}
  \begin{tabular}{cccccccc}
    \hline\hline
    & $\bret{\xi^1}_K$ & $\bret{\xi^2}_K$  & $\bret{\xi^1}^\parallel_{K^*}$ &
    $\bret{\xi^2}^\parallel_{K^*}$\rule[-7pt]{0pt}{20pt}\\\hline
    $24^3$ & 0.02893(87)(166)& 0.267(11)(13)& 0.0342(16)(21)& 0.248(17)(12) \\
    $16^3$ & 0.0277(17)(16)& 0.282(17)(14)& 0.0297(11)(16)& 0.255(13)(13)
    \\\hline
    \cite{Braun:2007zr}+\cite{Braun:2006dg} & 0.0272(5) & 0.260(6) & 0.033(2)(4) \\\hline\hline
    & $\bret{\xi^2}_\pi$ & $\bret{\xi^2}^\parallel_\rho$ &
    $\bret{\xi^2}^\parallel_\phi$\rule[-7pt]{0pt}{20pt}\\\hline
    $24^3$ & 0.272(15)(13)& 0.237(36)(12)& 0.246(10)(12) \\
    $16^3$ & 0.274(34)(13)& 0.245(27)(12)& 0.245(11)(12) \\\hline
    \cite{Braun:2006dg} & 0.269(39) \\\hline\hline
  \end{tabular}
\end{table}

\section{Non-perturbative renormalisation}

The previous section made use of perturbative renormalisation only. Since this
introduces uncertainties due to the known bad convergence of the perturbative
expansion, our task now is to compute the necessary renormalisation constants
non-perturbatively. For this, we use the Rome-Southampton method which employs a
simple renormalisation condition that is useful for any
regularisation \cite{Martinelli:1994ty}:
\begin{equation}
  \label{eq:rencond}
   \Lambda_\op{O}(p)
   = Z_\op{O}(\mu) Z^{-1}_q\, \Lambda^\text{bare}_\op{O}(p)\Bigr|_{p^2=\mu^2}
   = 1.
\end{equation}
Here $\Lambda^{(\text{bare})}_\op{O}(p)$ is the renormalised (bare) vertex
amplitude (definition in Eq.~\eqref{eq:amp+proj}), $Z_q$ is the quark field
renormalisation ($\psi = Z_q^{1/2} \psi^\text{bare}$) and $Z_\op{O}(\mu)$ the
desired renormalisation constant ($\op{O} = Z_\op{O} \op{O}^\text{bare}$) at the
scale $\mu$. The renormalisation condition should be applied for scales within
an appropriate window \cite{Martinelli:1994ty}, $\Lambda_\text{QCD} \ll \mu \ll
1/a$, that we know from our previous calculation \cite{Aoki:2007xm}.

To obtain the bare vertex function, we start by calculating the unamputated
Green's function of the operator $\op{O}$ between external off-shell quarks with
exceptional momenta, i.e.\ $p=p'$
\begin{equation}
  \label{eq:unampGreen}
  G_\op{O}(p) = \bret{\psi(p) \op{O}(0) \bar{\psi}(p)}
  = \sum_{x} \bret{
    \gamma_5 \left[ \textstyle{\sum_{y}} S(x,y) e^{\I
        py} \right]^\dagger \gamma_5\,
    J_\Gamma(x,x')\,
    \left[ \textstyle{\sum_{z}} S(x',z) e^{\I pz}\right] }.
\end{equation}
Where the operator is written as $\op{O}(q) = \sum_{x,x'} \exp(\I qx)
\bar{\psi}(x) J_\Gamma(x,x') \psi(x')$ with Dirac structure $\Gamma$ and
possible derivatives. We have already written the quark propagators and their
Fourier transform in Eq.~\eqref{eq:unampGreen} in a way suggesting the use of
momentum sources \cite{Gockeler:1998ye}. Instead of using a point source for the
inversion, one can perform the inversion on a momentum source $e^{\I pz}$ to
find a solution for $S(p)_x = \textstyle{\sum_{y}} S(x,y) e^{\I py}$. We then
amputate the Green's function and project onto the bare (tree-level) vertex
amplitude,
\begin{equation}
  \label{eq:amp+proj}
  \Pi_\op{O}(p) = \bret{S(p)}_G^{-1} \bret{G_\op{O}(p)}_G \bret{S(p)}_G^{-1}
  \quad\text{and}\quad
  \Lambda_\op{O}(p) = \frac{1}{N} \Tr \left( \Pi_\op{O}(p) P_\op{O} \right).
\end{equation}
Here the subscript $G$ indicates the gauge average and $N$ ensures the overall
normalisation of the trace. The projector $P_\op{O}$ has to match the specific
Lorentz and kinematical structure of the operator. Examples for quark bilinears
without derivatives can be found e.g.\ in \cite{Aoki:2007xm}. The operators
involving derivatives have a more complicated decomposition that is consistent
with their symmetries and tracelessness. Thus more care has to be taken to
project onto the correct tree-level contribution, see e.g.\ 
\cite{Gracey:2003mr}. One possibility for an operator like $\bar{\psi}
\gamma_{\{\mu} \lrd_{\nu\}} \psi$ would be
\begin{equation}
  \label{eq:projector}
  \Lambda_{\gamma_\mu D_\nu}(p) 
  = \frac{1}{6} \sum_{\substack{\mu,\nu\\\mu\le\nu}}
  \left[ \frac{\Tr \left[ \Pi_{\gamma_\mu D_\nu}(p)(\gamma_\mu+\gamma_\nu)
      \right]}{12(\hat{p}^\mu+\hat{p}^\nu)}
    -  \frac{{\textstyle{\sum_{\rho\neq\mu,\nu}}}\Tr \left[ \Pi_{\gamma_\mu
          D_\nu}(p) \gamma_\rho \right]}{12 \sum_{\rho\neq\mu,\nu}
      \hat{p}^\rho}\right].
\end{equation}
This particular example averages over all possible space-time components of the
operator and projector. The symmetrisation of the operator is reflected in the
first part of the projection, while the second part ensures we pick up the part
proportional to the tree-level contribution. We use $\hat{p}_\mu = \sin(p_\mu)$
to compensate the kinematic factors.

\begin{figure}[!t]
  \addtolength{\abovecaptionskip}{-6pt}
  \parbox[t]{.485\textwidth}{\centering\tiny\input{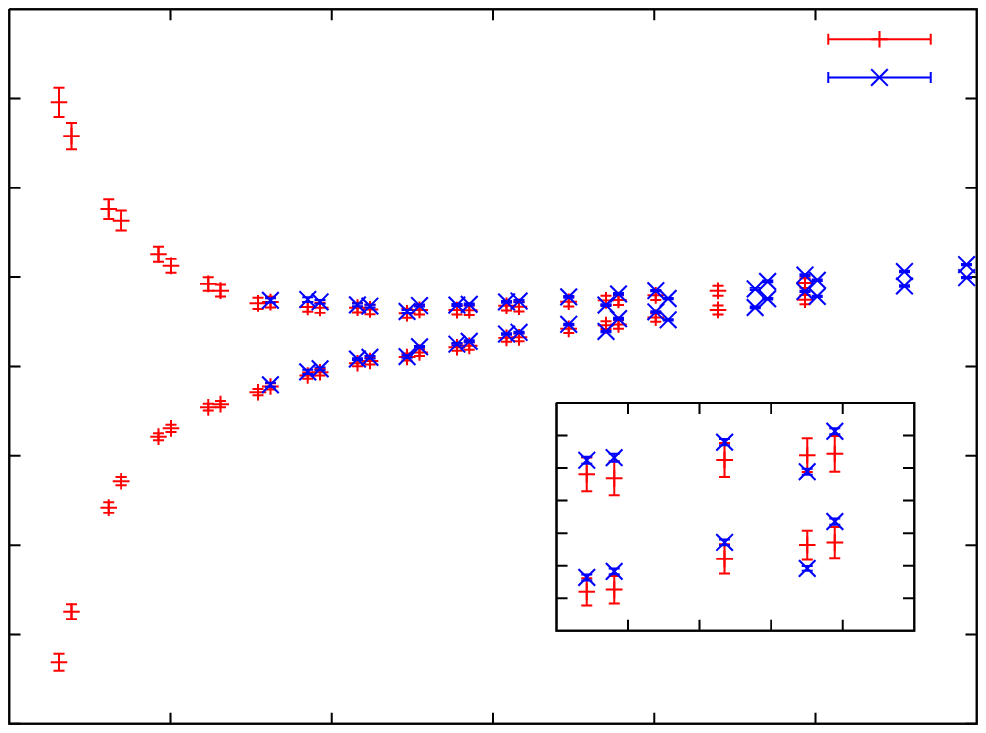}
    \caption{Comparison of the bare vertex amplitude for the vector and
      axial current from point and momentum sources. Both on $16^3\times 32$
      lattices for $am_\text{q} = 0.03$.\label{fig:comp1}}
  }\rule{.03\textwidth}{0pt}
  \parbox[t]{.485\textwidth}{\centering\tiny\input{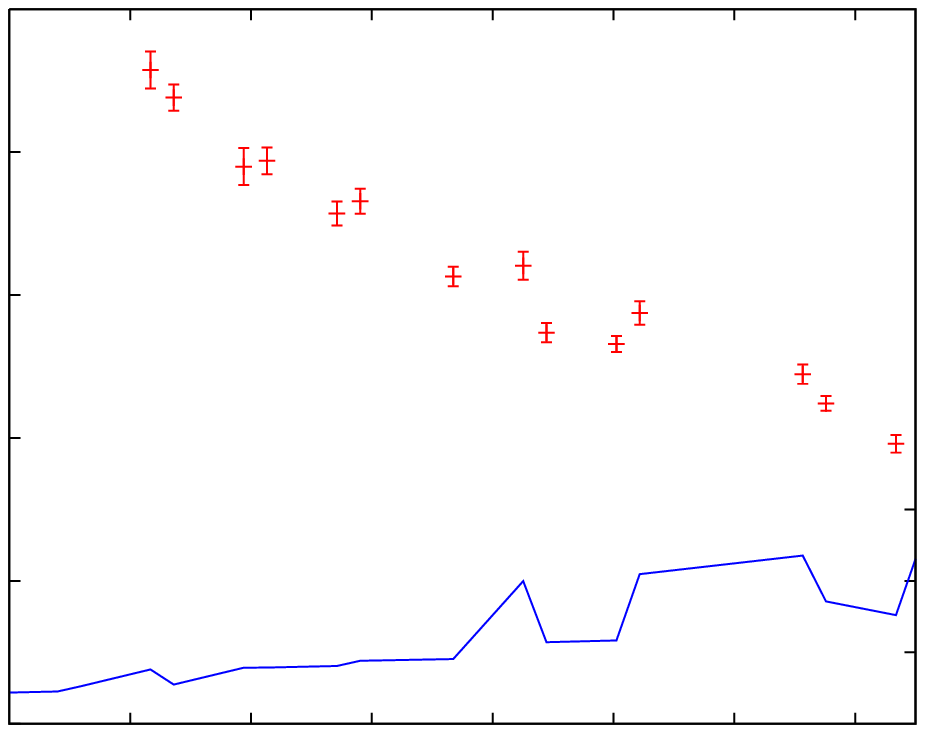}
    \caption{The bare vertex amplitude for the scalar density together with an
      indicator for discretisation errors, $\op{S}$ (see text). Again
      $16^3\times32$, $am_\text{q} = 0.03$.\label{fig:S}}
  }\\[2ex]
  \parbox[t]{.485\textwidth}{\centering\tiny\input{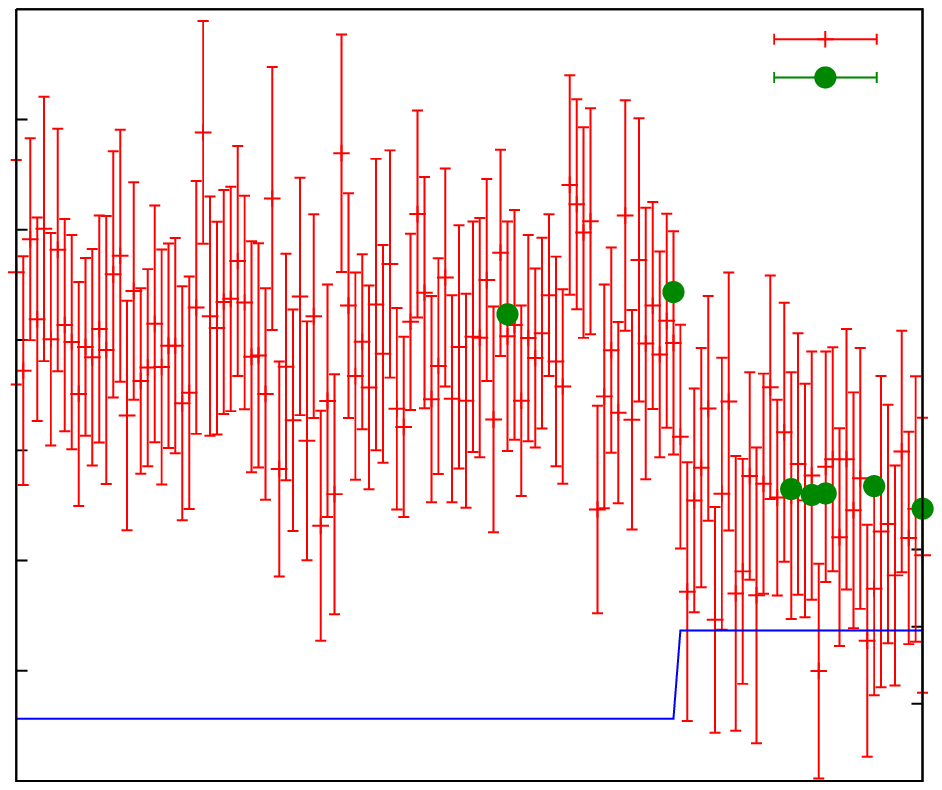}
    \caption{Normalised bare vertex for the two source types, again $16^3\times
      32$, $am_\text{q} = 0.03$. Only momentum sources reveal different
      discretisation errors.\label{fig:comp2}} 
  }\rule{.03\textwidth}{0pt}
  \parbox[t]{.485\textwidth}{\centering\tiny\input{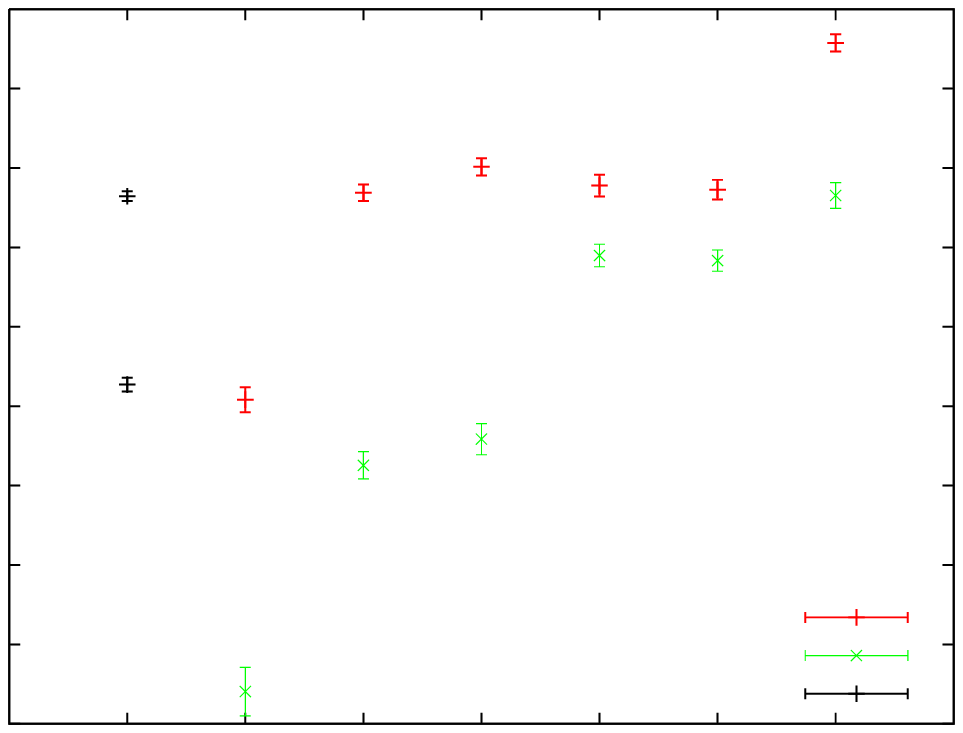}
    \caption{The vertex of the one derivative operator for two momenta, split up
      into the different choices of Lorentz indices $\{\mu\nu\}$. Again
      $16^3\times 32$.\label{fig:deriv}} }
\end{figure}

Since this is an ongoing project, let us only mention a few important findings
concerning the advantages of the momentum sources. Fig.~\ref{fig:comp1} compares
results from our earlier calculation using point sources, see
\cite{Aoki:2007xm}, with the current results. Of course the two calculations
agree, the important point is the much smaller statistical error for the
momentum sources as shown in the insert of Fig.~\ref{fig:comp1}. These results
have been obtained from one direction of the momentum for each $(ap)^2$ and up
to 25 configurations. The point source results on the other hand used 4 separate
sources with 75 configurations each and average multiple directions for one
value of $(ap)^2$, \cite{Aoki:2007xm}. It is thus possible to reduce the
computational cost while having smaller statistical errors.

The smaller errors make it possible to see lattice discretisation errors which
is shown in Figs.~\ref{fig:S}-\ref{fig:deriv}. Let us consider quark bilinears
without derivatives first. Here, one can imagine additional contributions from
terms $\propto p^2$. Using latticised momenta $\hat{p}$ again and expanding the
sine function, we expect possible discretisation errors to be proportional to
$\op{S} = \sum_\mu \left(2\pi p_\mu/L_\mu\right)^4$ (taking the leading
correction only). In Fig.~\ref{fig:S} we plot the bare vertex amplitude of the
scalar density along with $\op{S}$. The deviations from the expected smooth
behaviour of the vertex is clearly correlated with large changes in $\op{S}$.
This becomes even clearer, when we look at a fixed $(ap)^2$ but momenta in
different directions. The bare vertex should not depend on the latter. However,
the different directions result in different values of $\op{S}$.
Fig.~\ref{fig:comp2} shows again the scalar density, normalised with the mean
vector and axial vector vertex. Included are all directions to one momentum as
used in \cite{Aoki:2007xm}, sorted according to the corresponding value for
$\op{S}$.  The different discretisation errors are only visible for results
obtained with momentum sources however. Some examples of these are included in
the plot (note that their errors are hidden by the symbols).

Finally, let us mention that the discretisation errors are more difficult for
operators with derivatives. Here one can distinguish different directions of the
Lorentz indices of the operator w.r.t.\ the direction of the momentum. This fact
is demonstrated for two momenta and the vertex for $\bar{\psi} \gamma_{\{\mu}
\lrd_{\nu\}}\psi$ in Fig.~\ref{fig:deriv}. In the continuum all directions are
equivalent, on the lattice only choices which have similar momentum components
agree within errors, e.g.\ the combinations $\{\mu\nu\} = \{23\}, \{24\}$.

\section{Summary}

\vspace*{-.5ex}We have presented an update on our (preliminary) results for
distribution amplitudes which have been extended to more pseudo-scalar and
vector mesons. The $SU(3)$-breaking effects already found for the $K$ have been
confirmed for the $K^*$. We did not see a clear sign of finite size effects.

We have also presented our first findings using momentum sources for a
non-perturbative renormalisation of our lattice results. Here we see a clear
advantage in reducing the statistical errors making it possible to better
control the effects due to the lattice discretisation.

\section*{Acknowledgements}

\vspace*{-.5ex}We thank our colleagues in RBC and UKQCD within whose programme
this calculation was performed. We thank the QCDOC design team for developing
the QCDOC machine and its software. This development and the computers used in
this calculation were funded by the U.S.DOE grant DE-FG02-92ER40699, PPARC JIF
grant PPA/J/S/1998/0075620 and by RIKEN. This work was supported by DOE grant
DE-FG02-92ER40699, PPARC grants PPA/G/O/2002/00465 and PP/D000238/1, STFC Grant
PP/D000211/1 and from EU contract MRTN-CT-2006-035482 (Flavianet). We thank the
University of Edinburgh, PPARC, RIKEN, BNL and the U.S.\ DOE for providing the
QCDOC facilities used in this calculation.


\begin{thebibliography}{10}
  
\bibitem{Boyle:2006pw} {\bf UKQCD} Collaboration, P.~Boyle {\em et.~al.}, {\it
    {A lattice computation of the first moment of the kaon's distribution
      amplitude}}, {\em Phys. Lett.} {\bf B641} (2006) 67--74
  [\href{http://arXiv.org/abs/hep-lat/0607018}{{\tt hep-lat/0607018}}].
  
\bibitem{Donnellan:2007xr}{\bf UKQCD and RBC} Collaboration, M.~Donnellan {\em
    et.~al.}, {\it {Lattice Results for Vector Meson Couplings and Parton
      Distribution Amplitudes}}, \pos{PoS(LAT2007)369},
  [\href{http://arXiv.org/abs/0710.0869}{{\tt arXiv:0710.0869 [hep-lat]}}].
  
\bibitem{Martinelli:1994ty} G.~Martinelli {\em et.~al.}, {\it A general method
    for nonperturbative renormalization of lattice operators}, {\em Nucl. Phys.}
  {\bf B445} (1995) 81--108 [\href{http://arXiv.org/abs/hep-lat/9411010}{{\tt
      hep-lat/9411010}}].
  
\bibitem{Gockeler:1998ye} {\bf QCDSF} Collaboration, M.~G{\"o}ckeler {\em
    et.~al.}, {\it Non-perturbative renormalisation of composite operators in
    lattice {QCD}}, {\em Nucl. Phys.} {\bf B544} (1999) 699--733
  [\href{http://arXiv.org/abs/hep-lat/9807044}{{\tt hep-lat/9807044}}].

\bibitem{Allton:2007hx}
{\bf RBC and UKQCD} Collaboration, C.~Allton {\em et.~al.}, {\it {2+1 flavor
  domain wall QCD on a (2 fm)$^3$ lattice: light meson spectroscopy with $L_s =
  16$}},  {\em Phys. Rev.} {\bf D76} (2007) 014504
  [\href{http://arXiv.org/abs/hep-lat/0701013}{{\tt hep-lat/0701013}}].
  
\bibitem{Allton:2008pn} {\bf RBC and UKQCD} Collaboration, C.~Allton {\em
    et.~al.}, {\it {Physical Results from 2+1 Flavor Domain Wall QCD and SU(2)
      Chiral Perturbation Theory}}, \href{http://arXiv.org/abs/0804.0473}{{\tt
      arXiv:0804.0473 [hep-lat]}}.
  
\bibitem{Braun:2007zr} {\bf QCDSF and UKQCD} Collaboration, V.~Braun {\em
    et.~al.}, {\it {Distribution Amplitudes of Vector Mesons}},
  \pos{PoS(LAT2007)144}, [\href{http://arXiv.org/abs/0711.2174}{{\tt
      arXiv:0711.2174 [hep-lat]}}].

\bibitem{Chen:2003fp}
J.-W. Chen and I.~W. Stewart, {\it {Model independent results for SU(3)
  violation in light- cone distribution functions}},  {\em Phys. Rev. Lett.}
  {\bf 92} (2004) 202001 [\href{http://arXiv.org/abs/hep-ph/0311285}{{\tt
  hep-ph/0311285}}].
  
\bibitem{Braun:2006dg} {\bf QCDSF and UKQCD} Collaboration, V.~Braun {\em
    et.~al.}, {\it {Moments of pseudoscalar meson distribution amplitudes from
      the lattice}}, {\em Phys. Rev.} {\bf D74} (2006) 074501
  [\href{http://arXiv.org/abs/hep-lat/0606012}{{\tt hep-lat/0606012}}].
  
\bibitem{Aoki:2007xm}{\bf RBC and UKQCD} Collaboration, Y.~Aoki {\em et.~al.},
  {\it {Non-perturbative renormalization of quark bilinear operators and $B_K$
      using domain wall fermions}}, \href{http://arXiv.org/abs/arXiv:0712.1061
    [hep-lat]}{{\tt arXiv:0712.1061 [hep-lat]}}.

\bibitem{Gracey:2003mr}
J.~A. Gracey, {\it {Three loop anomalous dimension of the second moment of the
  transversity operator in the $\overline{MS}$ and {RI'} schemes}},  {\em Nucl.
  Phys.} {\bf B667} (2003) 242--260
  [\href{http://arXiv.org/abs/hep-ph/0306163}{{\tt hep-ph/0306163}}].

\end{thebibliography}
\providecommand{\href}[2]{#2}\begingroup\raggedright\endgroup

\end{document}